\documentclass[12pt]{article}
\usepackage{epsf}
\usepackage{graphicx}

\begin{document}

\title{ \bf Symmetry of the Neutron and Proton Superfluidity Effects
in Cooling Neutron Stars
}

\author{ {\bf M.E.\ Gusakov$^1$\footnote{{\it e-mail}:
gusakov@astro.ioffe.rssi.ru} $\,$,
A.D. Kaminker$^1$, D.G. Yakovlev$^1$, 
 O.Y. Gnedin$^2$} \\
     {\it $^1$  Ioffe Physical Technical Institute,} \\
     {\it Politekhnicheskaya 26, 194021 St.-Petersburg, Russia} \\
{\it $^2$ Space Telescope Science Institute,} \\
{ \it 3700 San Martin Drive, Baltimore, MD 21218, USA} \\ \\
{\rm Key words. stars: neutron -- dense matter}}
\date{${}$}
\maketitle
\def\la{\;\raise0.3ex\hbox{$<$\kern-0.75em\raise-1.1ex\hbox{$\sim$}}\;}
\def\ga{\;\raise0.3ex\hbox{$>$\kern-0.75em\raise-1.1ex\hbox{$\sim$}}\;}
\def\pFn{p_{\raise-0.3ex\hbox{{\scriptsize F$\!$\raise-0.03ex\hbox{\it n}}}} }  
\def\pFp{p_{\raise-0.3ex\hbox{{\scriptsize F$\!$\raise-0.03ex\hbox{\it p}}}} }  
\def\pFe{p_{\raise-0.3ex\hbox{{\scriptsize F$\!$\raise-0.03ex\hbox{\it e}}}} }  
\def\pFl{p_{\raise-0.3ex\hbox{{\scriptsize F$\!$\raise-0.03ex\hbox{\it l}}}} }  
\def\m@th{\mathsurround=0pt }
\def\eqalign#1{\null\,\vcenter{\openup1\jot \m@th
   \ialign{\strut$\displaystyle{##}$&$\displaystyle{{}##}$\hfil
   \crcr#1\crcr}}\,}
\newcommand{\vp}{\mbox{\boldmath $p$}}         
\newcommand{\vS}{\mbox{\boldmath $S$}}
\newcommand{\vP}{\mbox{\boldmath $P$}}
\newcommand{\om}{\mbox{$\omega$}}              
\newcommand{\Om}{\mbox{$\Omega$}}              
\newcommand{\Th}{\mbox{$\Theta$}}              
\newcommand{\ph}{\mbox{$\varphi$}}             
\newcommand{\del}{\mbox{$\delta$}}             
\newcommand{\Del}{\mbox{$\Delta$}}             
\newcommand{\lam}{\mbox{$\lambda$}}            
\newcommand{\Lam}{\mbox{$\Lambda$}}            
\newcommand{\ep}{\mbox{$\varepsilon$}}         
\newcommand{\ka}{\mbox{$\kappa$}}              
\newcommand{\dd}{\mbox{d}}                     
\newcommand{\vect}[1]{\bf #1}                
\newcommand{\vtr}[1]{\mbox{\boldmath $#1$}}  
\newcommand{\vF}{\mbox{$v_{\mbox{\raisebox{-0.3ex}{\scriptsize F}}}$}}  
\newcommand{\pF}{\mbox{$p_{\mbox{\raisebox{-0.3ex}{\scriptsize F}}}$}}  
\newcommand{\kF}{\mbox{$k_{\rm F}$}}           
\newcommand{\kTF}{\mbox{$k_{\rm TF}$}}         
\newcommand{\kB}{\mbox{$k_{\rm B}$}}           
\newcommand{\tn}{\mbox{$T_{{\rm c}n}$}}        
\newcommand{\tp}{\mbox{$T_{{\rm c}p}$}}        
\newcommand{\te}{\mbox{$T_{eff}$}}             
\newcommand{\ex}{\mbox{\rm e}}                 
\newcommand{\rate}{\mbox{$\frac{ \mbox{М╖░}}{\mbox{ка$^3 \cdot $к}}$}}
\newcommand{\mur}{\raisebox{0.2ex}{\mbox{\scriptsize (▄)}}} 
\newcommand{\Mn}{\raisebox{0.2ex}{\mbox{\scriptsize (▄{\it n\/})}}}        %
\newcommand{\Mp}{\raisebox{0.2ex}{\mbox{\scriptsize (▄{\it p\/})}}}        %
\newcommand{\MN}{\raisebox{0.2ex}{\mbox{\scriptsize (▄{\it N\/})}}}        %
\begin{abstract}
We investigate the combined effect of neutron 
and proton superfluidities on the cooling of
neutron stars whose cores consist of nucleons 
and electrons. 
We consider singlet-state pairing of protons
and triplet-state pairing of neutrons
in the cores of neutron stars. 
The critical superfluid temperatures $T_{\rm c}$ are
assumed to depend on the density of matter.
We study two types of neutron pairing 
with different components
of the total angular momentum of Cooper pairs
along the quantization axis ($|m_J|$ =0 or 2).
Our calculations are compared with observations
of thermal emission from isolated neutron stars.
We show that the observations can be interpreted 
by using two classes of superfluidity models: 
({\it 1}) strong proton superfluidity with 
a maximum critical temperature in the stellar core 
$T_{\rm c}^{\rm max} \ga 4 \times 10^9$~K
and weak neutron
superfluidity of any type 
($T_{\rm c}^{\rm max} \la 2 \times 10^8$~K);
 ({\it 2}) strong neutron superfluidity 
(pairing with $|m_J|$=0) and
weak proton superfluidity. 
The two types of models reflect an approximate 
symmetry with respect to an
interchange of the critical temperatures of neutron
and proton pairing.
\end{abstract}

\newpage

\section{Introduction}
\label{sect-1}
At present, the properties of superdense matter
in the cores of neutron stars are known poorly. For
example, the fundamental problem of the equation
of state for matter of supranuclear density has not yet
been solved. The existing calculations are model dependent
and yield a variety of equations of state
in the cores of neutron stars (Lattimer and
Prakash 2001; Haensel 2003) with different compositions
of this matter (nucleons, hyperons, pion or
kaon condensates, quarks). The properties of nucleon
superfluidity in the inner layers of neutron stars are
also unclear. Calculated critical temperatures of nucleon superfluidity
strongly depend on the nucleon-nucleon
interaction model used and on the method
of allowance for many-body effects (see, e.g., Lombardo
and Schulze 2001). However, they can be
studied by comparing the cooling theory of neutron stars with
observations of thermal emission from isolated
neutron stars.
Here, we continue to simulate the cooling of superfluid
neutron stars whose cores contain neutrons,
protons, and electrons, with critical temperatures
of nucleon superfluidity
depending on the density of matter.

We extend the class of cooling models that were proposed
by Kaminker et al.\ (2001, 2002) and Yakovlev
et al.\ (2001a, 2002) to interpret observations of
thermal emission from isolated neutron stars. These
authors paid special attention to the case of strong
proton superfluidity and weak neutron superfluidity in
the stellar core. Because superfluidity models have
a large uncertainty, we consider a broader class of
models without assuming from the outset that
proton pairing is stronger than neutron pairing.
In addition, attention is given to the model of nonstandard
neutron triplet-state pairing with an anisotropic gap
that vanishes along the quantization axis.

\section{Observational data}  
\label{sect-2}

The observational data on thermal emission from
eleven isolated middle-aged
($10^3 \la t \la 10^6$~yr)
neutron stars are collected in Table. In what follows,
$T_{\rm s}^\infty$ 
is the stellar surface temperature as detected by a
distant observer, and $t$ is the stellar age.
The data differ from those presented previously 
(see, e.g., Yakovlev et al.\ 2002), because they include the results
of new observations.

Two young objects, RX~J0822--4300 and
1E~1207.4--5209 (=J1210--5226), are radio-quiet
neutron stars in supernova remnants. Two of the
three oldest objects 
($t \ga 5 \times 10^5$~yr),  
RX~J1856.4--3754 and RX~J0720.4--3125, are also radio-quiet
neutron stars. The remaining seven sources ---
PSR~J0205+6449, the Crab pulsar (PSR~B0531+21),
the Vela pulsar (PSR~B0833--45), PSR~B1706--44,
PSR~J0538+2817, Geminga (PSR~B0633+1746),
and PSR~B1055--52 --- are observed as radio pulsars.

PSR~J0205+6449 and the Crab pulsar are located
in the remnants of historical supernovae; their ages
are known exactly. 

The age of RX~J0822--4300
was determined from the age of the remnant of the
host supernova Puppis A and lies within the range
$t=(2-5) \times 10^3$~yr
(see, e.g., Arendt et al.\ 1991),
with the most probable value  
$t = 3.7 \times 10^3$ yr
(Winkler et al.\ 1988).

The age of 1E~1207.4--5209 is
assumed to be equal to the age of the remnant of the
host supernova G296.5+10. According to Roger et
al.\ (1988), this age ranges from
$\sim 3 \times 10^3$~yr
to $\sim 20 \times 10^3$~yr. 

The age of the Vela pulsar is assumed
to lie within the range from the standard characteristic
pulsar age of 
$1.1 \times 10^4$~yr
to the age of 
$2.5 \times 10^4$~yr,
obtained by Lyne et al.\ (1996) by analyzing the pulsar
spindown with allowance for observed pulsar
glitches. 

Kramer et al.\ (2003) estimated the age of
PSR~J0538+2817, 
$t = (30 \pm 4)$ kyr,
from the
measured proper motion of the neutron star relative
to the center of the remnant of the host supernova
S147. 

The age of RX~J1856.4--3754 was estimated
by Walter (2001) from kinematic considerations and
revised by Walter and Lattimer (2002). Following the
latter authors, we take a mean value of 
$t=5 \times 10^5$ yr 
and choose an errorbar for $t$ that excludes the
value of 
$t = 9 \times 10^5$ yr 
obtained by Walter (2001).

Zane et al.\ (2002) and Kaplan et al.\ (2002) estimated
the characteristic age of RX~J0720.4--3125 from
X-ray measurements of the spindown rate of
the star $\dot{P}$. 
We take a mean value of $1.3 \times 10^6$ yr
with an uncertainty by a factor of 2.

The ages of the
three radio pulsars --- PSR~B1706--44, Geminga, and
PSR~B1055--52 --- are set equal to their characteristic
ages with the same uncertainty factor of 2.
%
\newcommand{\rrr}{\rule{0cm}{0.4cm}}
\newcommand{\hh}{\rule{0.5cm}{0cm}}
\newcommand{\hb}{\rule{0.4cm}{0cm}}
  
\begin{table}[!t]
\caption[]{
Surface temperatures of isolated neutron stars 
}
\label{tab-data}
\begin{center}
\begin{tabular}{|| l | c | c | c | c | l ||}
\hline
\hline
Source   & t   & $T_{\rm s}^\infty$   & Mo-         & Confi- & References \\
         & [$10^3$ yr] & [$10^6$~K]   & del $^{a)}$ & dence  &            \\ 
\hline
\hline
PSR~J0205+6449 
& 0.82 & $<$1.1 & bb  & --  & Slane et al.\ (2002) \\ 
%
Crab
& 1 & $<$2.0 & bb & 99.7$\%$  & Weisskopf et al.\ (2004) \\
%
RX~J0822--4300 &  2--5  &   1.6--1.9 & H  &  90\% & Zavlin et al.\ (1999) \\
1E 1207.4--5209  & 3--20 &  1.4--1.9 & H  & 90\%  & Zavlin et al.\ (2003) \\
Vela
& 11--25 & 0.65--0.71 & H &  68\%  & Pavlov et al.\ (2001) \\
PSR~B1706--44 & $\sim 17$ & $0.82^{+0.01}_{-0.34}$ & H & $68\%$ &
McGowan et al.\ (2004) \\
PSR~J0538+2817
& $30\pm4$ & $\sim 0.87$ & H & -- & Zavlin, Pavlov\ (2003) \\
Geminga
& $\sim 340$ & $\sim 0.5$  
& bb & 90\% & Zavlin, Pavlov\ (2003) \\
RX~J1856.4--3754 & $\sim 500$ & $<$0.65 & bb   
& -- & see text \\
PSR~B1055--52 &  $ \sim 540$ & $\sim 0.75$ 
& bb & -- & Pavlov, Zavlin\ (2003) \\
RX~J0720.4--3125 & $ \sim 1300$ & $\sim 0.51$ &
H   & -- & Motch et al.\ (2003) \\ 
\hline
\end{tabular}
\begin{tabular}{l}
$^{a)}\,$\rrr{\footnotesize 
Observations interpreted in terms of either
a hydrogen atmosphere model (H)}  \\[-0.5ex] 
\rrr{\footnotesize
$\; \; \;$ or the blackbody model (bb)
} \\[-0.5ex]
%
\end{tabular}
\end{center}
\end{table}  
%

For the two youngest objects (the Crab pulsar
and PSR~J0205+6449), only upper limits were
placed on $T^{\infty}_{\rm s}$ (Weisskopf et al.\ 2004;
Slane et al.\ 2002). The surface temperatures of five sources ---
RX~J0822--4300, 1E~1207.4--5209, Vela, PSR~B1706--44, 
and PSR~J0538+2817 --- were determined by
using neutron-star hydrogen atmosphere models (for
references, see Table).
These models give more
realistic neutron star radii and hydrogen column
densities (see, e.g., Pavlov et al.\ 2002) than the
blackbody model.

The pulsar PSR~B0656+14 that was considered
previously (see, e.g., Yakovlev et al.\ 2002) is excluded
from Table. A simultaneous analysis of
new X-ray and optical observations of the source
(with the improved distance to it obtained from the parallax
measurements by Brisken et al.\ 2003) leads either
to an overly small neutron-star radius (in the blackbody
model) or to an overly small distance to the
star (in the hydrogen atmosphere model); see
Zavlin and Pavlov (2003). This makes the interpretation of
thermal emission of the source too unreliable.

For Geminga and PSR~B1055--52, the blackbody
model is more self-consistent. Therefore, we take the
values of $T^{\infty}_{\rm s}$ obtained by interpreting the observed
spectra using this model. For PSR~B1055--52,
we take $T^{\infty}_{\rm s}$ from Pavlov and Zavlin (2003).

The surface temperature of RX~J1856.4--3754 has
not been determined accurately enough. The wide
spread in $T^{\infty}_{\rm s}$ obtained for different radiation models
(see, e.g., Pons et al.\ 2002; Braje and Romani 2002;
Burwitz et al.\ 2003; Pavlov and Zavlin 2003; Tr\"umper
et al.\ 2003) stems from the fact that optical and
X-ray observations cannot be described by a single
blackbody model. This may be attributed, for example,
to the presence of hot spots on the stellar
surface. Therefore, we fix only the upper limit of 
$T^{\infty}_{\rm s} < 6.5 \times 10^5$~K
that agrees with the value of $T^{\infty}_{\rm s}$ obtained
in the model of a Si-ash atmosphere (Pons et
al.\ 2002) and in the model of condensed matter on
the stellar surface (Burwitz et al.\ 2003). This limit is
consistent also with the model of a nonuniform stellar
surface temperature distribution proposed by Pavlov
and Zavlin (2003). For the latter model, the mean
stellar surface temperature is 
$T^{\infty}_{\rm s} = 5 \times 10^5$ K and
lies below the chosen upper limit.

Finally, we take the surface temperature of
RX~J0720.4--3125 from the paper by Motch et
al.\ (2003). These authors interpreted the observed
spectrum using an hydrogen atmosphere
model of finite depth.

For PSR~J0538--4300, PSR~B1055--52, and
RX~J0720--3125, the errors in $T^{\infty}_{\rm s}$ 
were not given
by the authors (see Table). In all these cases, we
assume them to be equal to 20$\%$.

\section{Models for nucleon superfluidity and neutrino emission
due to Cooper pairing of protons}
\label{sect-3}

Neutron or proton superfluidity can be characterized
by the critical temperature as a function of density,
$T_{\rm c}(\rho)$. Microscopic theories predict 
(see, e.g., Lombardo and Schulze 2001; for references, see also
the review by Yakovlev et al.\ 1999a) the existence
of singlet-state ($^1S_0$) neutron pairing
($T_{\rm cn} = T_{\rm cns}$) in the
inner crust and the outermost layers of the stellar
core and singlet-state proton pairing ($T_{\rm cp}$)
and triplet-state ($^3P_2$)
neutron pairing ($T_{\rm cn} = T_{\rm cnt}$) in the stellar core. 
One should bear in mind
the possibility of different components
$m_J$ of momentum of neutron-neutron pairs
with respect to the
quantization axis ($|m_J|$=0, 1, 2)
when triplet-state pairing is considered.
A superposition of states with different $m_J$ can also
be an energetically favored state of Cooper pairs
(see, e.g., Amundsen and Ostgaard 1985; Baldo et
al.\ 1992; Khodel et al.\ 1998, 2001). Only one type
of triplet-state superfluidity with $m_J = 0$ has commonly
been assumed in calculations of neutron star cooling
(except for the papers by Schaab et al.\ 1998, and Gusakov
and Gnedin 2002). Below,
we consider triplet-state pairing of neutrons with $|m_J|=$
0 and 2, because the effects of these two types of
superfluidity on the heat capacity and the neutrino
luminosity of neutron stars are qualitatively different.
Following Yakovlev et al.\ (1999a), we denote the
three types of superfluidity considered here --
$^1S_0$, $^3P_2$($m_J=0$), and
$^3P_2$($|m_J|=2$) --- by the letters A, B,
and C, respectively. The energy gap in the neutron
energy spectrum, $\epsilon({\bf p})$, is isotropic in case A and
anisotropic in cases B and C (i.e., it depends on the
angle between the particle momentum
{\bf p} and the quantization axis $z$). In case C, the
energy gap vanishes in the directions parallel and antiparallel to the
$z$ axis.

Nucleon superfluidity suppresses the neutrino
processes involving nucleons, changes the nucleon
heat capacity, and triggers an additional neutrino
emission mechanism related to Cooper
pairing of nucleons (Flowers et al.\ 1976). The
effect of neutron superfluidity C on the heat
capacity of the matter and the neutrino reactions
differs qualitatively from the effect of
superfluidity A or B. For example, the suppression
of the neutrino processes and the heat capacity by
superfluidity C and superfluidity B or A
has power-law and exponential character, respectively
(see, e.g., Yakovlev et al.\ 1999a).

Microscopic theories predict a variety of $T_{\rm c}(\rho)$
profiles (see, e.g., Lombardo and Schulze 2001). The peaks of
$T_{\rm c}(\rho)$ can take on
values from 
$\la 10^8$~K to $5 \times 10^{10}$~K. In many models,
the peaks of $T_{\rm cnt}(\rho)$ are lower
than the peaks of $T_{\rm cp}(\rho)$ and $T_{\rm cns}(\rho)$,
because of weaker nucleon-nucleon
attraction in triplet-state channels.

We use four phenomenological model
profiles $T_{\rm c}(\rho)$ of critical temperature
(for both, neutrons and protons) in the
core of a neutron star. In Fig.\ 1, these models are
denoted by {\it a, b, c,} and {\it d}. 
The chosen $T_{\rm c}(\rho)$ profiles
are similar and differ only in height (maximum value):
$T^{\rm max}_{\rm c} = 10^{10}$, 
$4.0 \times 10^9$, 
$8.0 \times 10^8$, 
and $8.0 \times 10^7$ K
(models {\it a, b, c,} and {\it d}). Superfluidities {\it a, b, c,} and
{\it d} will be called strong, moderately strong, moderate,
and weak, respectively. The chosen models are consistent
with theoretical calculations of $T_{\rm c}(\rho)$. The
$T_{\rm c}(\rho)$ curves have steep slopes at 
$\rho > \rho_{\rm D}$, 
where $\rho_{\rm D}$ is
the threshold density at which the direct Urca process
is open (see below).

\begin{figure}[ht]
\setlength{\unitlength}{1mm}
\leavevmode
\hskip  25mm
\includegraphics[width=100mm,bb=17  148  572  695,clip]{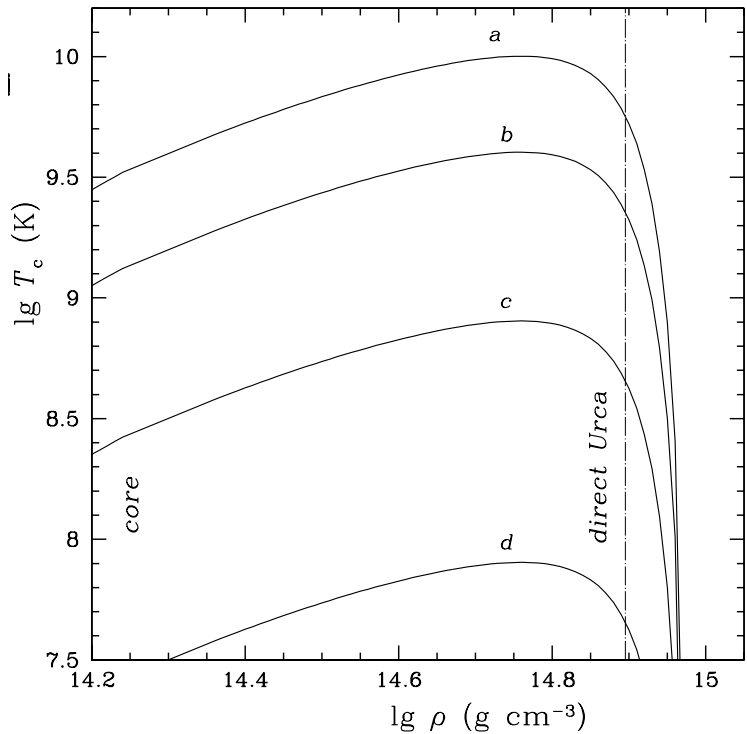}
\caption{
\footnotesize
Model density profiles of the critical neutron
and/or proton temperatures in the core of a neutron star.
The vertical dot-dashed line indicates the threshold density
at which the direct Urca process becomes operative.
}
\label{ModelSF}   
\end{figure}

Below, we will denote combined nucleon superfluidity
by $\alpha \beta$, where $\alpha$ 
is one of neutron triplet-state
(type B or C) superfluidity models ({\it a, b, c,} or {\it d}), 
and $\beta$ is
one of proton singlet-state superfluidity models ({\it a, b, c,}
or {\it d}).

Note a large uncertainty in the
neutrino emissivity $Q_{\rm p}$ due to
Cooper pairing of protons.
In the non-relativistic approximation (Yakovlev et
al.\ 1999b),
$Q_p \propto \zeta_p$, where $\zeta_p=c_{\rm Vp}^2$, and
$c_{\rm Vp} \approx 0.08$
is the vector constant of neutral current of protons;
the latter constant is numerically small and leads to unreasonably low
values of $Q_{\rm p}$. For comparison, the neutrino
emissivity $Q_{\rm n}$ due to
triplet-state pairing of neutrons
is proportional to 
$\zeta_n=c_{\rm Vn}^2+2c_{\rm An}^2=4.17$,
where $c_{\rm Vn}=1$ and $c_{\rm An}=-1.26$ 
are the vector and axial vector
constants of neutral neutron current, respectively.
According to Kaminker et al.\ (1999), the inclusion of
the relativistic correction that contains the axial vector
constant $c_{\rm Ap} = 1.26$ of neutron current of protons
can greatly (by a factor of 10 to 50) increase the
constant $\zeta_p$ (and the emissivity $Q_{\rm p}$)
compared to the non-relativistic value of
$\zeta_p=c_{\rm Vp}^2=0.0064$.

We used this value of $\zeta_p$, enhanced by relativistic
effects, in our previous simulations of neutron star cooling.
On the other hand, while studying the cooling
of stars with density dependent critical temperatures of protons
$T_{\rm cp}(\rho)$ (see, e.g., Kaminker et al.\ 2002),
we restricted ourselves to the models for strong proton superfluidity
(similar to model {\it a}). Such a superfluidity
arises at early cooling stages. At these stages, the neutrino emission
due to proton pairing cannot
compete with other neutrino processes and plays no
special role. In a cooler star, this neutrino emission
is generated only in a small volume and weakly affects
the cooling. Thus, in the cooling
scenarios with strong proton superfluidity, considered
previously, the emission due to Cooper pairing of protons
was unimportant (as well as the exact value of $\zeta_p$).

In this paper, we consider (among others) models of moderate
proton superfluidity in which the emission due
to proton pairing can appreciably affect the
cooling, so that the value of $\zeta_p$ is
important. 
As noted, for instance, by Yakovlev
et al.\ (1999b) and Kaminker et al.\ (1999), the
constant $\zeta_p$ can be affected not only by relativistic
effects, but also by renormalization
due to many-body effects in nucleon matter.
This renormalization for the process in question has
not yet been performed.
Carter and Prakash (2002) gave an
example of a similar renormalization of the constant of
the axial vector current. For the sake of definiteness,
we perform calculations by choosing the renormalized
value of $\zeta_p = 1$. The sensitivity of our calculations
to the value of $\zeta_p$ is described in Section 5.3 (also see Fig.\ 7).

\section{Cooling of stars with strong proton superfluidity}
\label{sect-4}

Let us compare the observational data with our
calculated cooling curves
($T^{\infty}_{\rm s} (t)$ profiles).
The calculations were performed
using a code described by Gnedin et al.\ (2001). As
in previous papers (mentioned in the introduction), we consider
the models of neutron stars whose cores are
composed of neutrons n, protons p, and electrons e.
We use a moderately stiff equation of state
in the stellar core proposed by Prakash et al.\ (1988)
(model I with the compression modulus of symmetric
nucleon matter at saturation $K$ = 240 MeV). The
maximum mass of a stable neutron star, for the chosen
equation of state, is $M = 1.977 M_{\odot}$
(at a radius of $R = 10.754$ km and a central density of 
$\rho_{\rm c} = 2.575 \times 10^{15}$ g~cm$^{-3}$). 
This equation of state opens the intense
direct Urca process of neutrino emission
(Lattimer et al.\ 1991) at densities $\rho$
above the threshold
density 
$\rho_{\rm D} = 7.851 \times 10^{14}$ g cm$^{-3}$, 
i.e., in stars
with masses $M > M_{\rm D} = 1.358 M_{\odot}$. The radius of a
star with the threshold mass $M_{\rm D}$ is $R = 12.98$ km.

The thermal evolution of a neutron star consists of
three stages:

({\it 1}) the stage of thermal relaxation of the inner
stellar layers ($t \la 100$ yr);

({\it 2}) the subsequent stage of neutrino cooling
($10^2 \la t \la 10^5$ yr) of a star with an isothermal core
via neutrino emission from the inner layers (mainly from the core);

({\it 3}) the final stage of photon cooling ($t \ga 10^5$ yr)
via photon emission from the stellar surface.

The cooling theory for non-superfluid stars cannot
explain the entire set of observational data (see, e.g.,
Kaminker et al.\ 2002). However, the theory can be
reconciled with the observations by taking into account
nucleon superfluidity. According
to Kaminker et al.\ (2001), it is sufficient to assume
the presence of strong proton superfluidity and weak
neutron superfluidity in stellar cores.

Figure 2 shows cooling curves for neutron
stars of different masses with weak neutron superfluidity
{\it d} and strong proton superfluidity {\it a}. Such weak
neutron superfluidity switches on only at the photon
cooling stage. Therefore, the type of weak neutron
superfluidity (B or C) does not affect the cooling of
middle-aged stars. The family of cooling curves for stars with
masses 
$M \ga M_{\odot}$ 
fill in the hatched region. All of the
observed sources fall within this region; i.e., they can
be interpreted in terms of the assumed superfluidity
model.

As shown Kaminker et al.\ (2002), strong proton
superfluidity (with weak neutron superfluidity or with
normal neutrons) gives rise to three types of cooling
neutron stars.

Low-mass stars cool down very slowly (more
slowly than low-mass non-superfluid stars). Cooling
curves for such stars weakly depend on their
mass, the equation of state in their cores,
and proton superfluidity model (on the specific
form of the $T_{\rm cp}(\rho)$ profile as long as superfluidity
in the stellar core is strong enough, 
$T_{\rm cp}(\rho) \ga 4 \times 10^9$~K). 
The upper boundary of the hatched region
in Fig.\ 2 is the cooling curve for a star with a mass of
$M = 1.35 M_\odot$;
it is almost indistinguishable from the
cooling curve for a star with
$M = 1.1 M_{\odot}$ 
and agrees with the observations of four sources,
RX~J0822--4300, 1E~1207.4--5209, PSR~B1055--52,
and RX~J0720.4--3125, the hottest ones for their
ages. These sources will be considered as low-mass
neutron stars.

High-mass neutron stars cool down very rapidly
via intense neutrino emission generated by the direct
Urca process in the inner stellar core. At high
densities ($\rho \ga 10^{15}$ g cm$^{-3}$), 
proton superfluidity weakens (Fig.\ 1)
and ceases to suppress the
neutrino emission. The cooling curves for such stars
weakly depend on their mass, the equation of state, 
and the proton superfluidity model. They
almost coincide with the cooling curves for high-mass
non-superfluid stars. All of the observed isolated
neutron stars are much hotter than the stars of this
type.

Finally, medium-mass stars cool down moderately
rapidly. Their cooling strongly depends on
the mass, the equation of state, and
proton superfluidity model. By varying the stellar
mass, we can obtain a family of cooling curves
that fill the space between the cooling curves for
low-mass and high-mass stars. We consider the
sources PSR~J0205+64, Vela, PSR~B1706--44,
PSR~J05438+2817, Geminga, and 
RX~J1856.4--3754 as medium-mass neutron stars.
%
\begin{figure}[ht]
\setlength{\unitlength}{1mm}
\leavevmode
\hskip  25mm
\includegraphics[width=100mm,bb=18  145  573  698,clip]{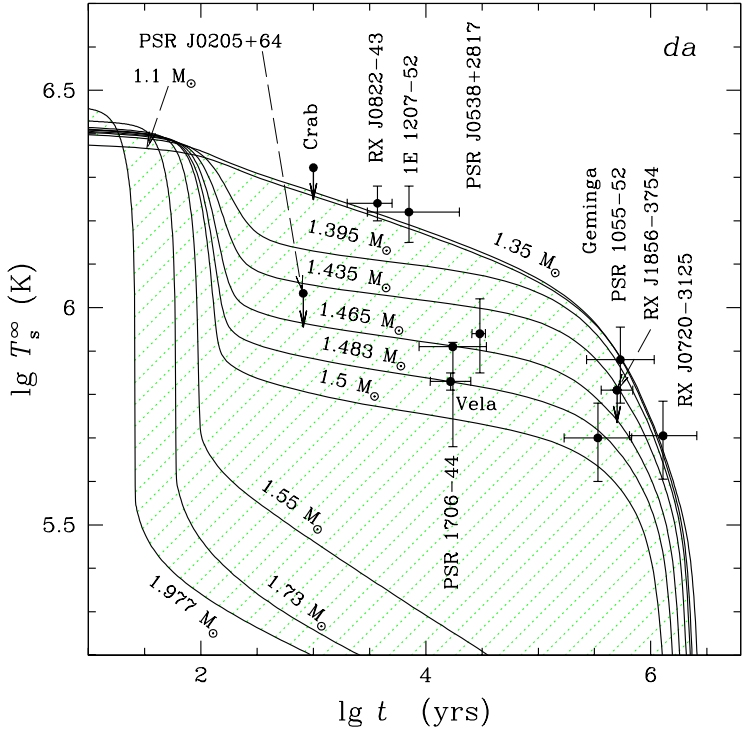}
\caption{
\footnotesize
Comparison of the observations (see Table)
with the cooling curves for neutron stars with masses
from 1.1 to $1.977 M_{\odot}$ 
(indicated near the curves) for weak
neutron superfluidity {\it d} and strong proton superfluidity {\it a}.
The region filled by cooling curves for stars of different
masses is hatched.
}
\label{ProtonSF}   
\end{figure}
%

\section{Cooling of neutron stars with combined
nucleon superfluidity}
\label{sect-5}

Figures 3--6 show cooling curves for neutron
stars with different superfluidities of neutrons $\alpha$ and
protons $\beta$ 
($\alpha$, $\beta$ = {\it a, b, c,} or {\it d}). 
Neutron superfluidity is of type B.
We consider all the possible
combinations of neutron and proton superfluidities.
For each combination
$\alpha \beta$, we show
the upper cooling curve of a low-mass star
($M = 1.1 M_{\odot}$, 
with the central density
$\rho_{\rm c} = 6.23 \times 10^{14}$ g cm${^-3}$) 
and the lower cooling curve of a high-mass
star ($M = M_{\rm max}$).
The lower curve is virtually independent of
the models of superfluidity $\alpha \beta$ 
(see the previous section).
The region between the upper and lower
curves (similar to the hatched region in Fig.\ 2) can
be filled by cooling curves of medium-mass
stars and is accessible by an assumed model of superfluidity
$\alpha \beta$.
As in Fig.\ 2, we show also the observational
data.
Nucleon superfluidity models
can be constrained by comparing accessible
$T^{\infty}_{\rm s}$ regions with the observational data.

Each of Figs.\ 3--6 consists of two panels: In
panel (a) neutron superfluidity $\alpha$ is fixed, and
cooling curves are given for all four proton superfluidity
models; in panel (b) proton superfluidity $\beta$
is fixed, and cooling curves are given for all
four neutron superfluidity models. By comparing
panels (a) and (b), we can trace the change in cooling
when proton superfluidity is replaced by
neutron superfluidity (and vice versa).

\begin{figure}[ht]
\setlength{\unitlength}{1mm}
\leavevmode
\hskip  0mm
\includegraphics[width=160mm,bb=18  145  544  419,clip]{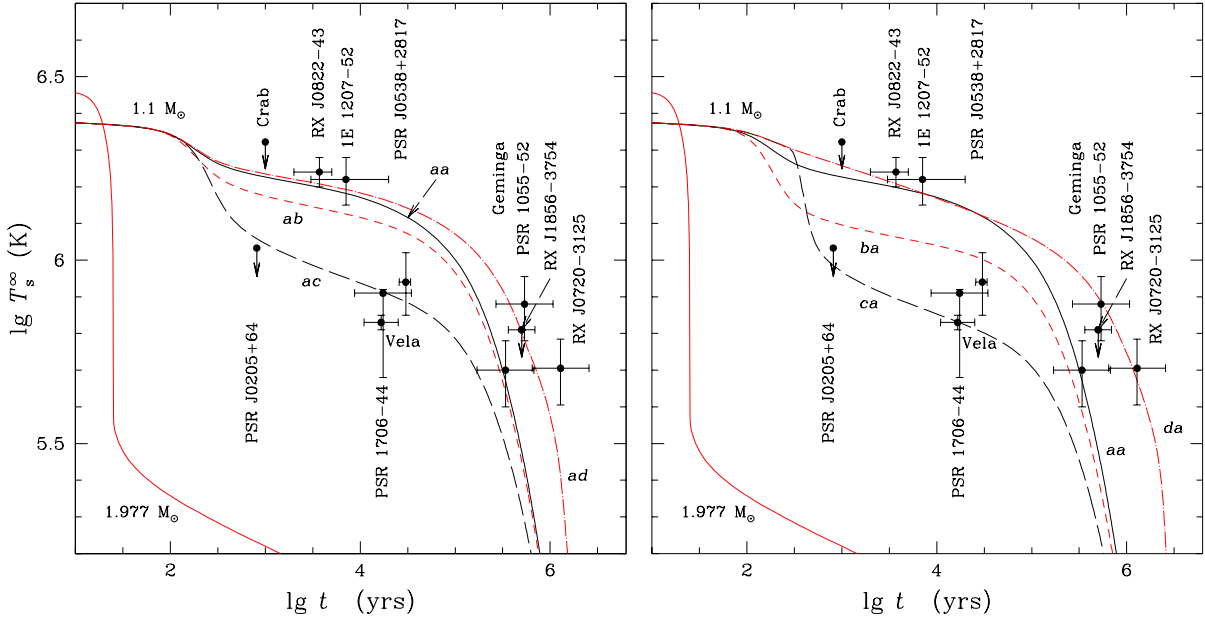}
\caption
{
(a) Cooling of stars of two masses, 
$M = 1.977 M_{\odot}$ and $M = 1.1 M_{\odot}$, 
for neutron superfluidity {\it a} and
different proton superfluidity models ({\it a, b, c,} or {\it d}); 
(b) the same for proton superfluidity model {\it a}
and different neutron
superfluidity models ({\it a, b, c,} or {\it d}). 
Neutron superfluidity B is employed in all the cases.
Theoretical curves are compared
with the observations. The cooling of a star with 
$M = 1.977 M_{\odot}$ does not depend on superfluidity model.
}
\label{33}   
\end{figure}
%
%
\begin{figure}[ht]
\setlength{\unitlength}{1mm}
\leavevmode
\hskip  0mm
\includegraphics[width=160mm,bb=18  145  544  419,clip]{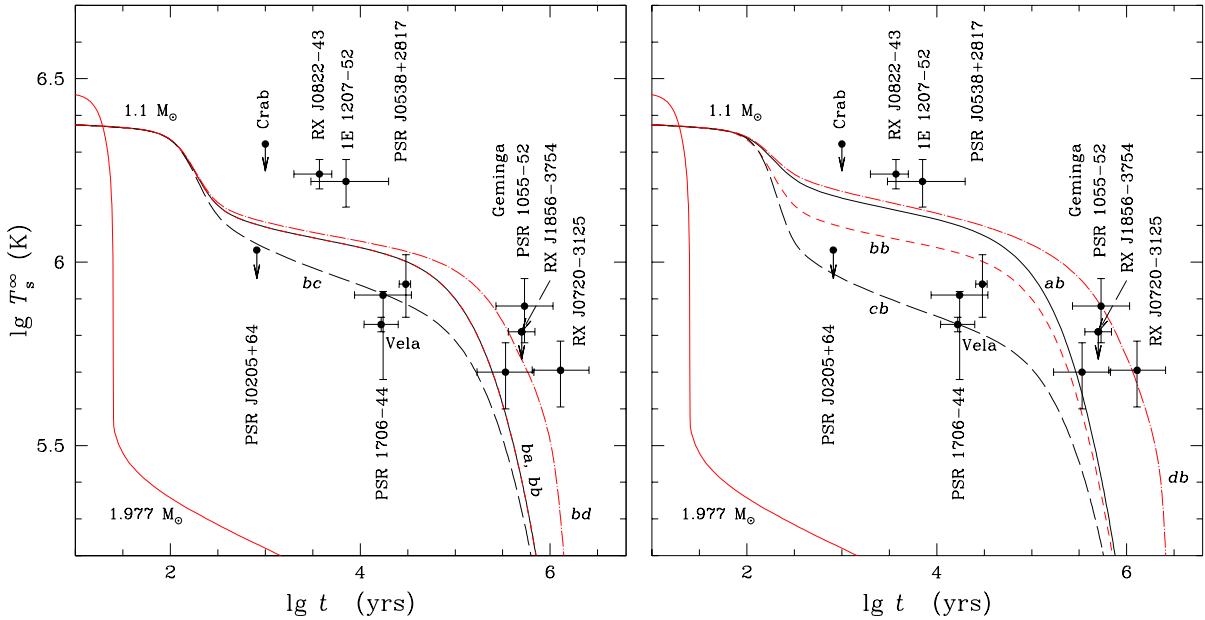}
\caption
{
Same as in Fig.\ 3 but for fixed superfluidity model {\it b}
of neutrons (a) or protons (b).
}
\label{44}   
\end{figure}

\subsection{Fixed proton superfluidity}
\label{fixed-protons}

Let us choose a proton superfluidity model ($\beta$ =
{\it a, b, c,} or {\it d}) and consider the dependence of the upper
cooling curves on the models of neutron superfluidity
$\alpha$ in Figs.\ 3--6.

The cooling curves $b \beta$ run below the
curves $a \beta$ because
of the neutrino emission due to Cooper
pairing of neutrons. This emission is significantly suppressed in
the models with strong neutron superfluidity {\it a} (see,
e.g., Yakovlev et al.\ 1999a, 1999b, 2001b). All
other neutrino reactions involving neutrons and
the neutron heat capacity are fully suppressed
by neutron superfluidity {\it a} or {\it b}.
The difference
between the cooling curves $a \beta$ and $b \beta$ depends on the
model for proton superfluidity $\beta$. Thus, for example,
as we go from the model $\beta = a$ to $\beta = b$ and then
to the model of moderate proton superfluidity $\beta = c$,
the contribution of the neutrino emission due to
Cooper pairing of protons to the neutrino luminosity of
the star increases (and becomes dominant for $\beta = c$).
Indeed, the neutrino emission due to Cooper pairing
affects most strongly the cooling at moderate critical
temperatures of nucleons, $T_{\rm c} \sim 2 \times (10^8-10^9)$~K
(see, e.g., Yakovlev et al.\ 1999b, 2001b). As a result, the
difference between the cooling curves $a \beta$ and $b \beta$ in
Figs.\ 3--5 steadily decreases as we go from the model
$\beta = a$ to $\beta = b$ and $\beta = c$. At the same time, the
accessible theoretical cooling regions disagree with the
observations more and more.

For weak superfluidity $\beta = d$ (Fig.\ 6), protons
remain normal as long as $t \la 10^5$ yr, until the onset of the
photon cooling stage. In this case, the main neutrino
process involving protons is the bremsstrahlung
in proton-proton collisions. Since the contribution
of this process to the neutrino emission is much
smaller than the contribution of Cooper
pairing of protons in model {\it c},
the difference between the curves {\it ad} and
{\it bd} again increases. As in Fig.\ 3 (curves {\it aa} and {\it ba}),
it is mainly determined by the more intense
neutrino generation due to Cooper pairing of neutrons in
model {\it b} than in model {\it a}. As a result, the accessible
region of stellar surface temperatures for combined
superfluidity {\it ad} (as for superfluidity {\it da},
cf.\ Fig.\ 6 with Figs.\ 2 and 3) agrees with the observations.

The curves $a \beta$ and $b \beta$ approach one another at
the photon cooling stage ($t \ga 10^5$ yr). In this case,
the influence of neutron superfluidity {\it a} or {\it b} on the
cooling of the star manifests itself mainly in strong
suppression of the neutron heat capacity. As a result,
the heat capacity of the star is determined by the
heat capacity of protons (also suppressed by
superfluidity $\beta$) and electrons.

For superfluidity models $c \beta$, the neutrino emission
due to Cooper pairing of neutrons is especially
efficient. Accordingly, the cooling curves $c \beta$ in
Figs.\ 3--6 run well below the curves $a \beta$ and $b \beta$ and do not
differ too much from one another. In particular, all of
the curves $c \beta$ describe the sharp speedup in cooling
at $t \sim 300$ yr associated with the switch-on of the neutrino
emission due to neutron pairing. We can see that the
accessible surface temperatures obtained for
the superfluidity models $c \beta$ lie well below most of the
observational data points.

For the model of weak neutron superfluidity {\it d}, the
$d \beta$ curves almost coincide with the cooling curves for
normal neutrons. Differences arise only at the photon
cooling stage ($t \ga 10^5$ yr) from the partial suppression
of the neutron heat capacity. However, at the
neutrino cooling stage, neutron superfluidity {\it d}
has not yet set in. Therefore, all of the $d \beta$ cooling
curves lie above the $c \beta$ curves. At $t \ga 10^5-10^6$ yr, the
$d \beta$ cooling curves for any $\beta$ run above the
curves $a \beta$ and $b \beta$
due to the strong suppression of the neutron
heat capacity by superfluidities {\it a} and {\it b}. Finally,
the cooling curve for the model of combined superfluidity
{\it dd} is close to the standard cooling curve for
non-superfluid low-mass ($M < M_{\rm D}$) neutron stars.
This cooling curve disagrees with the observations of
many neutron stars (both the hottest and coolest ones
for their ages).

\begin{figure}[ht]
\setlength{\unitlength}{1mm}
\leavevmode
\hskip  0mm
\includegraphics[width=160mm,bb=18  145  544  419,clip]{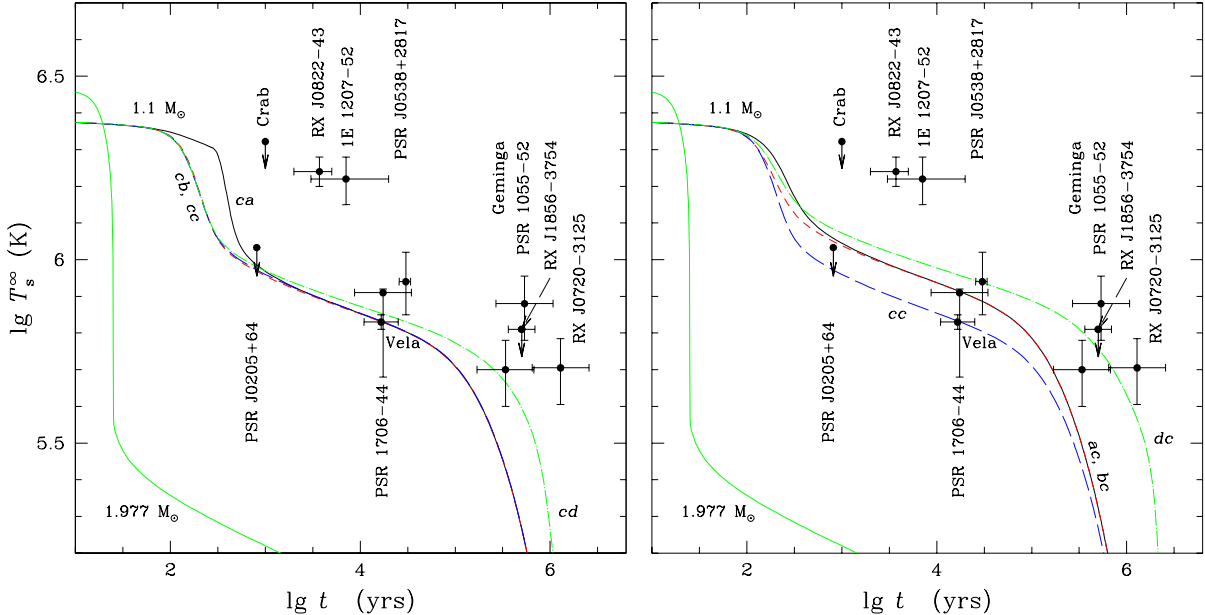}
\caption
{
Same as in Fig.\ 3 for superfluidity model {\it c}
of neutrons (a)
or protons (b).
}
\label{55}   
\end{figure}

\begin{figure}[ht]
\setlength{\unitlength}{1mm}
\leavevmode
\hskip  0mm
\includegraphics[width=160mm,bb=18  145  544  419,clip]{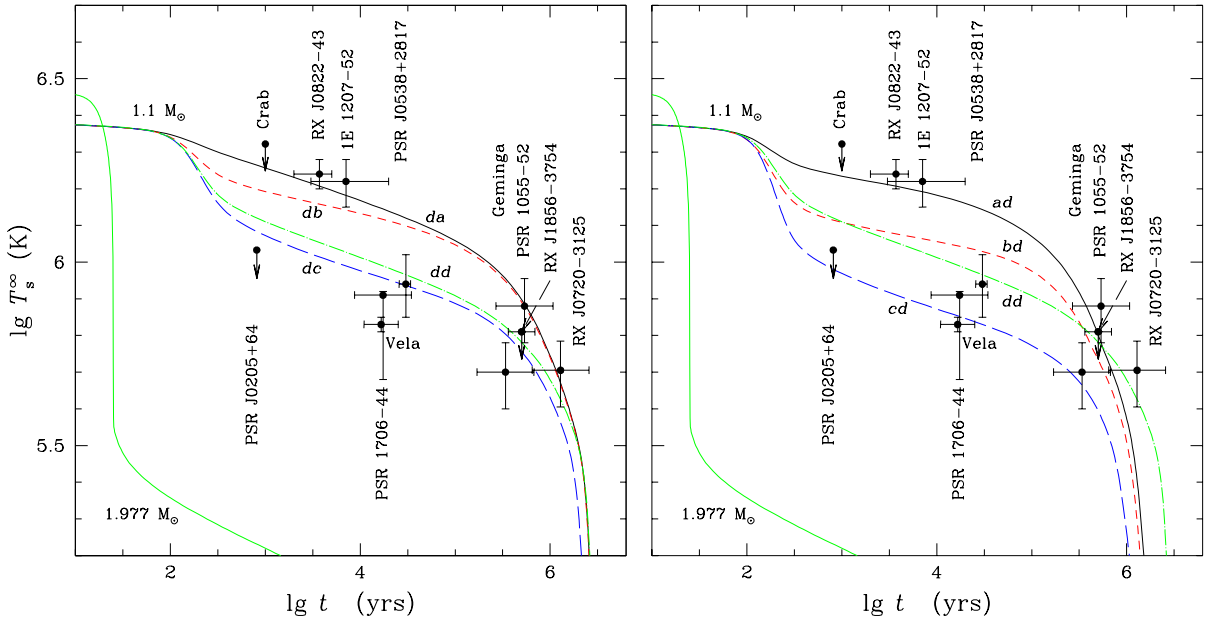}
\caption
{
Same as in Fig.\ 3 for superfluidity model {\it d}
of neutrons (a) or protons (b).
}
\label{66}   
\end{figure}

\subsection{Fixed Neutron Superfluidity}
\label{fixed-neutrons}

Let us choose a neutron superfluidity model ($\alpha =
a, b, c,$ or {\it d}) and consider the dependence of the upper
cooling curves on proton superfluidity models $\beta$. A
comparison of panels (a) and (b) in Figs.\ 3--6 reveals
a qualitative similarity between cooling curves of
low-mass ($M = 1.1 M_{\odot}$) stars where neutron and
proton superfluidities are inverted (i.e., for models
$\alpha \beta$ and $\beta \alpha$).

The quantitative differences between cooling curves
$\alpha \beta$ and $\beta \alpha$
are attributable to different neutron
and proton number densities in the cores of neutron
stars and to different types of neutron (triplet-state) and
proton (singlet-state) pairing. This results in a slightly
asymmetric effect of neutrons and protons on the
neutrino luminosity and the heat capacity (see, e.g.,
Yakovlev et al.\ 1999a). Thus, for example, at temperatures
$T$ slightly below $T_{\rm c}$, the neutrino emissivity
due to neutron pairing is approximately
an order of magnitude higher than that due to proton
pairing (even for the chosen constant $\zeta_p = 1$).
Therefore, the
cooling curves
$b( \beta = a, d)$ and $c(\beta = a, b, d)$
(panel (a) in Figs.\ 4 and 5) lie below the ``inverted''  cooling curves
$(\alpha = a, d) b$ and $(\alpha = a, b, d) c$
(panel (b) in the same figures). On the other hand,
the curve {\it ad} (Fig.\ 3a) at the photon cooling stage
($t \ga 10^5$ yr) runs below the curve {\it da} (Fig.\ 3b). This
is because the heat capacity of the neutron star core
is stronger suppressed by neutron superfluidity
$\alpha = a$ than by proton superfluidity $\beta = a$.
In other cases, the inversion of neutron
and proton superfluidities leads to qualitatively similar
(roughly symmetric) cooling curves in Figs.\ 3--6. For
high-mass neutron stars ($M > M_{\rm D}$), this symmetry
was found by Levenfish et al.\ (1999) in their simplified
cooling calculations for stars with constant critical temperatures of
neutrons and protons over the stellar core.

A comparison of the upper cooling curves with
the observations in Figs.\ 3--6 shows that there are
only two models of combined nucleon superfluidity
that are consistent with the set of observational data.
They include the model {\it da} discussed in Section 4
and the ``inverted'' model {\it ad} (Figs.\ 3 and 6). In other
words, one (neutron or proton) superfluidity must be
weak, while the other must be strong. Other
models are unable to simultaneously explain the
observational data, primarily for four neutron stars
(RX~J0822--4300, 1E~1207.4--5209, PSR~B1055--52, 
and RX~J0720.4--3125), 
the hottest ones for their ages.

Varying nucleon superfluidity models, we can
constrain critical temperatures of nucleons at which
the theory agrees with the observations. In general,
the following conditions must be satisfied simultaneously:
either 
$T_{\rm cnt}^{\rm max} \la 2 \times 10^8$~K and
$T_{\rm cp}^{\rm max} \ga 4 \times 10^9$~K,
or
$T_{\rm cnt}^{\rm max} \ga 5 \times 10^9$~K and
$T_{\rm cp}^{\rm max} \la 2 \times 10^8$~K.

The models of {\it moderate neutron and/or proton
superfluidity} in the cores of neutron stars with peak
temperatures 
$T^{\rm max}_{\rm cnt}$ and/or $T^{\rm max}_{\rm cp}$
in the range
$\sim (2 \times 10^8 - 4 \times 10^9)$~K
are inconsistent with the observations
of neutron stars hottest for their ages.
We can show that this conclusion is valid for a much
broader class of nucleon superfluidity models than
those used here. Nevertheless, there is a
narrow region of nucleon superfluidity parameters at
which the combination of strong nucleon superfluidity
of one type and moderate nucleon superfluidity of
another type can be reconciled with the observations
(for details, see Gusakov et al.\ 2004).

\begin{figure}[ht]
\setlength{\unitlength}{1mm}
\leavevmode
\hskip  0mm
\begin{center}
\includegraphics[width=80mm,bb=18  145  565  695,clip]{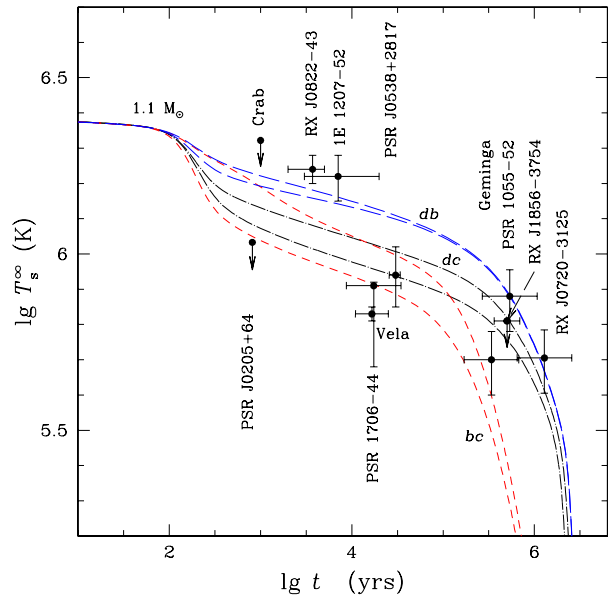}
\end{center}
\caption
{
Cooling curves of a low-mass star ($M = 1.1 M_{\odot}$) 
for three models of nucleon superfluidity ({\it db} -- long dashes,
{\it dc} -- dash-and-dot curves, {\it bc} -- short dashes) in comparison
with the observations. Lower and upper curves for
each superfluidity model are computed,
respectively, with the renormalized
($\zeta_p$ = 1) and non-renormalized constant in the expression
for neutrino emissivity due to Cooper pairing of protons.
}
\label{77}   
\end{figure}

\subsection{On the constant in the expression for the neutrino emissivity
due to Cooper pairing of protons}
\label{cooper}

Let us briefly discuss the sensitivity of cooling
curves to the constant $\zeta_p$ in the expression for the
neutrino emissivity due to proton
pairing. Recall that the value of $\zeta_p$ that includes many-body
effects is known poorly. In our calculations,
we used the (renormalized) value $\zeta_p = 1$.

For example, Fig.\ 7 shows cooling curves of
a low-mass neutron star for three models of neutron and proton
superfluidity ({\it db, dc,} and {\it bc}). As shown
above, the neutrino emission due to Cooper pairing
is especially important in low-mass stars.
As everywhere in this section, we consider
neutron superfluidity of type B. The lower of two curves
for each superfluidity model is calculated with the
renormalized constant $\zeta_p = 1$, while the upper curve
is calculated with the non-renormalized constant
(but taking into account relativistic effects;
see Kaminker et al.\ 1999).

In model {\it db}, proton superfluidity {\it b} is moderately
strong and appears at an early cooling stage.
The neutrino emission due to proton pairing plays a
relatively minor role, and the exact value of $\zeta_p$ weakly
affects the cooling.

In model {\it dc} and, especially, in model {\it bc}, moderate
proton pairing $\beta = c$ results in intense neutrino emission
and appreciably speeds up the cooling. In these 
cases, the cooling curves are most sensitive to $\zeta_p$.
However, as seen from Fig. 7,
the employed variations of $\zeta_p$
cannot lead to agreement of cooling curves {\it dc} and
{\it bc} with the observations and, hence,
do not affect our conclusions. We believe the renormalized
value $\zeta_p = 1$ to be more realistic than the
non-renormalized value. The existing uncertainty in $\zeta_p$
introduces an uncertainty in the cooling theory. In
particular, for the non-renormalized value of $\zeta_p$, the
approximate symmetry of cooling curves relative
to the inversion of nucleon superfluidity models
($\alpha \beta \rightleftharpoons \beta \alpha$) noted 
above is much less pronounced than
for the renormalized value (see also Yakovlev et
al.\ 1999a). It is possible that the
choice of $\zeta_p$ may
be important in the future for reconciling theory with
observations.

\section{Two types of triplet-state pairing of neutrons}
\label{sect-6}

Let us compare the effect of the two types of
neutron triplet-state superfluidity (B and C) on cooling
of neutron stars. Clearly, significant differences might
be expected for strong neutron superfluidity. As follows
from the results of the previous section, strong
neutron superfluidity of type B (model {\it a}) and weak
proton superfluidity (model {\it d}) can ensure agreement
between the theory and the observations. Let us
consider this case in more detail. Figure 8 shows the
cooling curves for neutron stars of different masses
with superfluidity {\it ad}: In panels (a) and (b), we take
neutron pairing of type B and C, respectively.

According to Fig.\ 8a
(as for model {\it da} in Fig.\ 2), we can identify
the same three types of cooling neutron stars: low-mass,
slowly cooling stars; high-mass, rapidly cooling
stars; and medium-mass stars with a moderate
cooling rate.

The cooling of neutron stars with neutron
superfluidity of type C was first calculated by Schaab et
al.\ (1998). However, these authors used an oversimplified
description of the effects of superfluidity on
neutrino reactions. More accurate calculations
were performed by Gusakov and Gnedin (2002), who
compared the results obtained for
superfluidities of types B and C. The authors used the approximation
of critical temperatures $T_{\rm cp}$
and $T_{\rm cnt}$ constant over the
stellar core. Calculations indicate that, 
in many cases,
cooling curves do not change if
the actual $T_{\rm cp}(\rho)$ and $T_{\rm cnt}(\rho)$
profiles are replaced by
effective
constant critical temperatures close to $T_{\rm cp}(\rho_{\rm c})$
and $T_{\rm cnt}(\rho_{\rm c})$ at the stellar center 
($\rho = \rho_{\rm c}$).
This approximation is valid if $T_{\rm c}(\rho)$
is a smooth function of $\rho$ near the stellar center 
(e.g., in low-mass stars).

Gusakov and Gnedin (2002) showed that
neutron superfluidity of type C speeds up the neutron star
cooling (compared to superfluidity of type B). This is
caused by the power-law suppression of the neutrino
emissivity by superfluidity C (in contrast
to the exponential suppression in case B; see, e.g.,
Yakovlev et al.\ 1999a; Gusakov 2002).
Our calculations (Fig.\ 8) indicate that the above conclusion
remains valid in a more realistic approach, taking
into account variations of the critical temperatures
$T_{\rm cnt}(\rho)$ and $T_{\rm cp}(\rho)$ over the stellar core.

Cooling curves for low- and medium-mass
stars in Fig.\ 8b lie well below cooling curves for stars of the
same masses in Fig.\ 8a. On the other hand,
cooling curves for high-mass stars ($M \ga 1.55 M_{\odot}$) 
in both panels of the figure almost coincide for the
obvious reason: the critical temperatures are low in
the central regions of these stars ($T_{\rm cnt} \la 10^8$~K, see
Fig.\ 1), so that superfluidity ceases to affect the cooling.

Thus, according to Fig.\ 8, strong neutron
superfluidity of type C disagrees with the observations of the
neutron stars hottest for their ages. This superfluidity
is still too weak to completely suppress the modified
Urca process in a low-mass star, thereby making
the star hotter. Of course, the theory can be reconciled
with the observations by choosing the model of
stronger superfluidity C. Our calculations indicate
that this requires a $T_{\rm cnt}(\rho)$ profile with the maximum
$T^{\rm max}_{\rm cnt} \sim 10^{11}$~K.
However, such a strong triplet-state
pairing seems unrealistic. 
Khodel et al.\ (1998, 2001) gave theoretical
arguments against the appearance
of superfluidity of type C in neutron stars.

\begin{figure}[ht]
\setlength{\unitlength}{1mm}
\leavevmode
\hskip  0mm
\includegraphics[width=160mm,bb=18  145  544  419,clip]{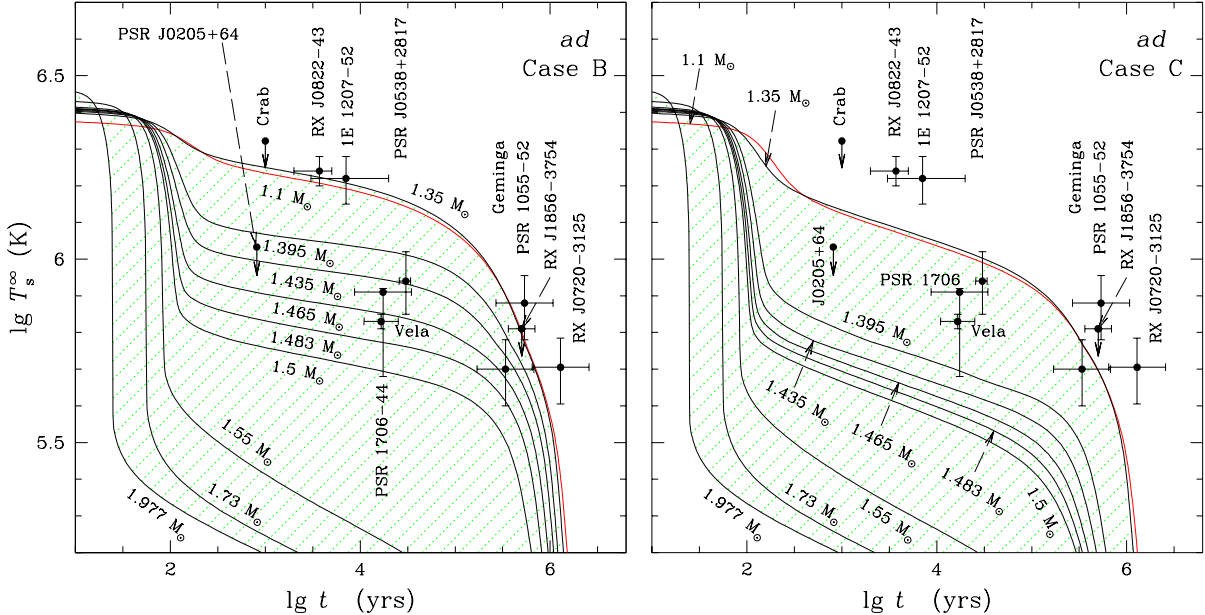}
\caption{
Same as in Fig.\ 2 for model {\it a} of strong
neutron superfluidity of type B (a)
or C (b) and model {\it d} of weak proton
superfluidity (of type A).
}
\label{88}   
\end{figure}

\section{Conclusions}
\label{sect-7}

We have analyzed the cooling of neutron stars with
triplet-state pairing of neutrons
and singlet-state pairing of protons
in stellar cores. Our results are summarized
below.

(1) Cooling curves are qualitatively
symmetric relative to the inversion of
neutron and proton superfluidity models. For low-mass
stars ($M < M_{\rm D}$), this symmetry is largely determined
by the dimensionless constant $\zeta_p \sim 1$ in the
expression for the neutrino emissivity due
to Cooper pairing of protons. At $\zeta_p \ll 1$, obtained without
the renormalization of $\zeta_p$ with account for many-body
effects, the symmetry is much less pronounced than
at $\zeta_p \sim 1$ expected after the renormalization.

(2) Two types of neutron and proton superfluidity
models are consistent with the observations of thermal
emission from isolated neutron stars. First, they
are the models of strong (type A) proton superfluidity
and weak (or absent) (type B or C)
neutron superfluidity with maximum (over the stellar
core) critical temperatures 
$T^{\rm max}_{\rm cp} \ga 4 \times 10^9$~K
and
$T_{\rm cnt}^{\rm max} \la 2 \times 10^8$~K. 
Second, they are the models
of strong (type B) neutron superfluidity and weak (or
absent) proton superfluidity with
$T_{\rm cnt}^{\rm max} \ga 5 \times 10^9$~K and
$T_{\rm cp}^{\rm max} \la 2 \times 10^8$~K.
The models of the
first type seem more realistic. Note, in particular, a
recent paper by Schwenk and Friman (2004) who
predicted a weakening of triplet-state pairing of neutrons
by many-body effects.

(3) Models of moderate (type B or C) neutron
superfluidity and/or moderate (type A) proton superfluidity
with maximum critical temperatures $T^{\rm max}_{\rm cnt}$
and $T^{\rm max}_{\rm cp}$
in the range from $\sim 2 \times 10^8$ to $\sim 4 \times 10^9$~K
are inconsistent with the observations of primarily the
hottest young neutron stars RX~J0822--4300 and
1E~1207.4--5209. However, the agreement between
the cooling theory and the observations is also
possible for a narrow range of parameters of
moderate neutron superfluidity
($T^{\rm max}_{\rm cnt} \sim 6 \times 10^8$ K). 
This possibility was considered by Gusakov et al.\ (2004).

(4) Strong triplet-state neutron pairing of type C
can appreciably speed up the cooling of middle-aged
($10^3 \la t \la 10^5$ yr) neutron stars compared
to pairing of type B
for the same $T_{\rm cnt}(\rho)$ profiles. For strong
neutron pairing of type C
($T^{\rm max}_{\rm cnt} \ga 5 \times 10^9$~K)
and weak proton pairing
($T^{\rm max}_{\rm  cp} \la 2 \times 10^8$~K),
the theory cannot be reconciled with the
observations of RX~J0822--4300, 1E~1207.4--5209,
PSR~B1055--52, and RX~J0720.4--3125, the hottest
sources for their ages. 
For any models of strong or
moderate neutron pairing, the transition from
pairing of type B to pairing of type C just enhances the
difference between the theory and the observations.

Our analysis is simplified, because we have considered
only the cores of neutron stars composed of neutrons,
protons, and electrons (disregarding the possible existence
of hyperons, pion and kaon condensates, or
quark matter). Moreover, we have chosen only one equation
of state in the cores of neutron stars
and similar profiles of the critical
temperatures $T_{\rm c}(\rho)$ of neutrons and protons
in the stellar cores. Varying the
equation of state (for the matter composed of neutrons, protons and
electrons) leads to shifting the threshold of
the direct Urca process (to changing $\rho_{\rm D}$ and $M_{\rm D}$).
Varying $T_{\rm c}(\rho)$ profiles (but retaining
their general shape) at high $T^{\rm max}_{\rm c} \ga 2 \times 10^9$~K
leads to shifting
the characteristic values of $\rho$ at which superfluidity
weakens and ceases to suppress the intense
neutrino emission. Both effects shift boundary
masses that separate the three types of cooling neutron
stars (Kaminker et al.\ 2002), but do not alter our
main conclusions. It is significant that the simplest model of
neutron stars with strong proton superfluidity (even
without neutron superfluidity) is capable of explaining
the available observations.

It should be noted that the cooling of neutron stars
also depends on  (singlet-state) neutron pairing
in the inner stellar crust, on the magnetic field in
the outermost stellar layers, and on the presence or
absence of a surface layer of light elements (see, e.g.,
Potekhin et al.\ 2003). In general, however, these factors
have a weaker effect on the cooling than nucleon
superfluidity in the cores of neutron stars considered
here. We have disregarded them by restricting our
analysis to superfluidity effects in the stellar cores.
Our cooling code allows us to include these factors,
and it can be necessary for
interpreting individual sources, primarily the
objects hottest for their ages
(see, e.g., Potekhin et al.\ 2003).
The cooling of neutron stars can also depend on
internal reheating mechanisms related, for example, to
the viscous dissipation of differential rotation
(see, e.g., Page 1998a, 1998b). We emphasize that
these reheating mechanisms are model dependent. On
the other hand, the available observations can be
interpreted without invoking them.

Note that the surface temperatures of neutron
stars $T^{\infty}_{\rm s}$ are difficult to determine from observational
data (see, e.g., Pavlov et al.\ 2002).
For this purpose, one needs reliable observational
data and theoretical models
of neutron star atmospheres.
The existing values of $T^{\infty}_{\rm s}$
can change appreciably,
which can strongly affect the interpretation
of the observations, especially, of RX~J0822--4300,
1E~1207.4--5209, PSR~B1055--52, and RX~J0720.4--3125. 
Future observations of thermal emission from
isolated neutron stars will be crucial for understanding
superfluid properties of dense matter in stellar
cores.

\begin{center}
{\bf \Large Acknowledgments}
\end{center}

This work was supported by the Russian Foundation
for Basic Research (project nos.~02-02-17668
and~03-07-90200), the Russian Leading Scientific
Schools Program (project no.~1115.2003.2), and the
INTAS~YSF~(grant~no.~03-55-2397).
\newpage


\noindent
\begin{center}
{\Large \bf  References}
\end{center}

\begin{list}{}{}

\item
L.\ Amundsen, E.\ {\O}stgaard,
Nucl.\ Phys.\ {\bf A 442}, 163 (1985).  

\item
R.G.\ Arendt, E.\ Dwek, R.\ Petre,
Astrophys.\ J. {\bf 368}, 474  (1991).  

\item
M.\ Baldo, J.\ Cugnon, A.\ Lejeune, U.\ Lombardo,
Nucl.\ Phys.\ {\bf A 536}, 349  (1992).

\item
T.M.\ Braje, R.W.\ Romani,
Astrophys.\ J.\ {\bf 580}, 1043 (2002).

\item
W.F.\ Brisken, S.E.\ Thorsett, A.\ Golden,
W.M.\ Goss,
Astrophys.\ J.\ Lett.\ {\bf 593}, L89 (2003).

\item
V.\ Burwitz, F.\ Haberl, R.\ Neuh\"auser,
P.\ Predehl, J.\ Tr\"umper, V.E.\ Zavlin,
Astron.\ Astrophys. {\bf 399}, 1109  (2003).

\item
G.W.\ Carter, M.\ Prakash,
Phys.\ Lett.\ {\bf B525}, 249 (2002).

\item
E.G.\ Flowers, M.\ Ruderman, P.G.\ Sutherland,
Astrophys.\ J.\ {\bf 205}, 541 (1976).  

\item
O.Y.\ Gnedin, D.G.\ Yakovlev, A.Y.\ Potekhin,
MNRAS {\bf 324}, 725 (2001). 

\item
M.E.\ Gusakov, Astron.\ Astrophys.\ 
{\bf 389}, 702 (2002).  

\item
M.E.\ Gusakov and O.Yu.\ Gnedin,
Astron.\ Lett.\ {\bf 28}, 669 (2002).

\item
M.E.\ Gusakov, A.D.\ Kaminker, D.G.\ Yakovlev, O.Yu.\ Gnedin,
Astron.\ Astrophys.\ {\bf  421}, 1143 (2004).

\item 
P.\ Haensel,
in: {\it Final Stages of Stellar Evolution} 
(Ed.\ J.-M.\ Hameury, C.\ Motch, EAS Publications Series, 
EDP Sciences, 2003), p.\ 249.

\item
A.D.\ Kaminker, P.\ Haensel, D.G.\ Yakovlev,
Astron.\ Astrophys.\ {\bf 345}, L14 (1999).

\item
A.D.\ Kaminker, P.\ Haensel, D.G.\ Yakovlev,
Astron.\ Astrophys.\ {\bf 373}, L17 (2001).  

\item
A.D.\ Kaminker, D.G.\ Yakovlev, O.Y.\ Gnedin,
Astron.\ Astrophys.\ {\bf 383}, 1076 (2002).  

\item
D.L.\ Kaplan, S.R.\ Kulkarni, M.H.\ van Kerkwijk,
H.L.\ Marshall, Astrophys.\ J.\ {\bf 570}, L79 (2002). 

\item
V.A.\ Khodel, V.V.\ Khodel, J.W.\ Clark,
Phys.\ Rev.\ Lett.\ {\bf 81}, 3828 (1998). 

\item
V.A.\ Khodel, J.W.\ Clark, M.V.\ Zverev,
Phys.\ Rev.\ Lett.\ {\bf 88}, 031103 (2001). 

\item
M.\ Kramer, A.G.\ Lyne, G.\ Hobbs, 
O.\ L\"ohmer, P.\ Carr, C.\ Jordan, A.\ Wolszczan,   
Astrophys.\ J.\ Lett.\ {\bf 593}, L31 (2003).

\item
J.M.\ Lattimer, M.\ Prakash,
Astrophys.\ J.\ {\bf 550}, 426 (2001). 

\item
J.M.\ Lattimer, C.J.\ Pethick, M.\ Prakash, P.\ Haensel,
Phys.\ Rev.\ Lett.\ {\bf 66}, 2701 (1991).

\item
K.P.\ Levenfish, Yu.A.\ Shibanov, D.G.\ Yakovlev,
Astron. Lett.\ {\bf 25}, 417 (1999).

\item
U.\ Lombardo, H.-J.\ Schulze,
in: {\it Physics of Neutron Star Interiors}
(Ed.\ D.\ Blaschke, N.K.\ Glendenning, A.\ Sedrakian;
Springer, Berlin, 2001) p.\ 30.

\item
A.G.\ Lyne, R.S.\ Pritchard, F.\ Graham-Smith, 
F.\ Camilo, Nature {\bf 381}, 497 (1996).

\item
K.E.\ McGowan, S.\ Zane, M.\ Cropper, J.A.\ Kennea,
F.A.\ Cordova, C.\ Ho, T.\ Sasseen, W.T.\ Vestrand, 
Astrophys.\ J.\ {\bf 600}, 343 (2004).

\item
C.\ Motch, V.E.\ Zavlin, F.\ Haberl,
Astron.\ Astrophys.\ {\bf 408}, 323 (2003).

\item
D.\ Page,
in: {\it The Many Faces of Neutron Stars},
NATO ASI Ser. C, v. 515,
(Ed.\ R.\ Buccheri, J.\ van Paradijs, M.A.\ Alpar;
Kluwer, Dordrecht, 1998a) p.\ 539. 

\item
D.\ Page,
in: {\it Neutron Stars and Pulsars}
(Ed.\ N.\ Shibazaki, N.\ Kawai, S.\ Shibata, T.\ Kifune;
Univ.\ Acad.\ Press, Tokyo, 1998b) p.\ 183.

\item
G.G.\ Pavlov, V.E.\ Zavlin,
in: {\it the Proceedings of the
XXI Texas Symposium on Relativistic Astrophysics}, 
(Ed.\ R.\ Bandiera, R.\ Maiolino, F.\ Mannucci;
World Scientific Publishing: Singapore, 2003) p.\ 319.

\item
G.G.\ Pavlov, V.E.\ Zavlin, D.\ Sanwal, V.\ Burwitz,
G.P.\ Garmire, 
Astrophys.\ J.\ Lett.\ {\bf 552}, L129 (2001).

\item
G.G.\ Pavlov, V.E.\ Zavlin, D.\ Sanwal,
in: {\it WE-Heraeus Seminar on Neutron Stars, Pulsars and Supernova Remnants No.270}
(Ed.\ W.\ Becker, H.\ Lesch, J.\ Tr\"umper; 
Garching: MPE-Report 278, 2002), p.\ 273.

\item
J.A.\ Pons, F.\ Walter, J.\ Lattimer, M.\ Prakash,
R.\ Neuh\"auser, P.\ An, 
Astrophys.\ J.\ {\bf 564}, 981 (2002).  

\item
A.Y.\ Potekhin, D.G.\ Yakovlev, G.\ Chabrier,
O.Y. Gnedin,
Astrophys.\ J.\ {\bf 594}, 404 (2003).

\item
M.\ Prakash, T.L.\ Ainsworth, J.M.\ Lattimer,
Phys.\ Rev.\ Lett.\ {\bf 61}, 2518 (1988).

\item
R.S.\ Roger, D.K.\ Milne, M.J.\ Kesteven,
K.J.\ Wellington, R.F.\ Haynes, 
Astrophys.\ J.\ {\bf 332}, 940 (1988).

\item
Ch.\ Schaab, F.\ Weber, M.K.\ Weigel,
Astron.\ Astrophys.\ {\bf 335}, 596 (1998).

\item
A.\ Schwenk, B.\ Friman,
Phys.\ Rev.\ Lett.\ {\bf 92}, 082501 (2004).

\item
P.O.\ Slane, D.J.\ Helfand, S.S.\ Murray,
Astrophys.\ J.\ Lett.\ {\bf 571}, L45 (2002).

\item
J.E.\ Tr\"umper, V.\ Burwitz, F.\ Haberl,
V.E. Zavlin, astro-ph/0312600 (2003).

\item
F.M.\ Walter, Astrophys.\ J.\ {\bf 549}, 433 (2001).

\item
F.M.\ Walter, J.M.\ Lattimer,
Astrophys.\ J.\ Lett.\ {\bf 576}, L145 (2002). 

\item
M.C.\ Weisskopf, S.L.\ O'Dell, F.\ Paerels, 
R.F.\ Elsner, W.\ Becker, A.F.\ Tennant, D.A.\ Swartz,
Astrophys.\ J.\ {\bf 601}, 1050 (2004).

\item
P.F.\ Winkler, J.H.\ Tuttle, R.P.\ Kirshner, M.J.\ Irwin,
in: {\it Supernova Remnants and the Interstellar Medium}
(Ed.\ R.S.\ Roger, T.L.\ Landecker; Cambridge: Cambridge University Press,
1988), p.\ 65. 

\item
D.G.\ Yakovlev, K.P.\ Levenfish , Yu.A.\ Shibanov,
Usp.\ Fiz.\ Nauk {\bf 169}, 825 (1999a).

\item
D.G.\ Yakovlev, A.D.\ Kaminker, K.P.\ Levenfish,
Astron.\ Astrophys.\ {\bf 343}, 650 (1999b).

\item
D.G.\ Yakovlev, A.D.\ Kaminker, O.Y.\ Gnedin,
Astron.\ Astrophys.\ {\bf 379}, L5 (2001a).

\item
D.G.\ Yakovlev, A.D.\ Kaminker, O.Y.\ Gnedin, 
P.\ Haensel, Phys.\ Rep.\ {\bf 354}, 1 (2001b).

\item
D.G.\ Yakovlev, O.Y.\ Gnedin, A.D.\ Kaminker,
A.Y.\ Potekhin, in:
{\it WE-Heraeus Seminar
on Neutron Stars, Pulsars and Supernova Remnants No. 270}
(Ed.\ W.\ Becker, H.\ Lesch, J.\ Tr\"umper; 
Garching: MPE-Report 278, 2002), p.\ 287.

\item
S.\ Zane, F.\ Haberl, M.\ Cropper, V.E.\ Zavlin,
D.\ Lumb, S.\ Sembay, C.\ Motch,
MNRAS {\bf 334}, 345 (2002).  

\item
V.E.\ Zavlin, G.G.\ Pavlov, 
Memorie della Societa' Astronomica Italiana,
the Proceedings of the EPIC Consortium
(held on Oct. 14--16, 2003 in Palermo)
to be published [astro-ph/0312326].

\item
V.E.\ Zavlin, J.\ Tr\"umper, G.G.\ Pavlov,
Astrophys.\ J.\ {\bf 525}, 959  (1999).

\item
V.E.\ Zavlin, G.G.\ Pavlov, D.\ Sanwal,
Astrophys.\ J. {\bf 606}, 444 (2004).

\end{list}

\end{document}